\documentclass[aps,prd,groupedaddress,showpacs,showkeys]{revtex4}
\usepackage{latexsym,epsfig,amsmath,amssymb}
\usepackage{amssymb}
\usepackage{amsmath}
\usepackage{amsfonts}
\usepackage{epsfig} 
\usepackage{verbatim}
\usepackage{bm}
\usepackage{mathtext}

\newcommand{\seq}{\begin{subequations}}
\newcommand{\sen}{\end{subequations}}
\newcommand{\eq}{\begin{eqnarray}}
\newcommand{\en}{\end{eqnarray}}
\newcommand{\ra}{\rangle}

\def\shiftdown#1{#1\llap{\lower.04ex\hbox{#1}}}

\def\sc{\Sigma_c}
\def\lc{\Lambda_c}

\def\L2{\Lambda^2}

\begin{document}

\title{Charmed baryon $\Sigma_c(2800)$ as a $ND$ hadronic molecule} 

\noindent
\author{
        Yubing Dong$^{1,2}$, 
        Amand  Faessler$^3$,   
        Thomas Gutsche$^3$, 
        Valery E. Lyubovitskij$^3$\footnote{On leave of absence
        from Department of Physics, Tomsk State University,
        634050 Tomsk, Russia}
\vspace*{1.2\baselineskip}}
\affiliation{
$^1$ Institute of High Energy Physics, Beijing 100049, P. R. China\\ 
\vspace*{.4\baselineskip} \\
$^2$ Theoretical Physics Center for Science Facilities (TPCSF), CAS, 
Beijing 100049, P. R. China\\ 
\vspace*{.4\baselineskip} \\ 
$^3$ Institut f\"ur Theoretische Physik,  Universit\"at T\"ubingen,\\
Kepler Center for Astro and Particle Physics, \\ 
Auf der Morgenstelle 14, D--72076 T\"ubingen, Germany\\} 

\date{\today}

\begin{abstract} 

The isotriplet $\sc (2800)$ baryon with possible quantum numbers
$J^{\rm P} =  \frac{1}{2}^{\pm}$ or $\frac{3}{2}^{\pm}$ is considered 
as a hadronic molecule composed of a nucleon and a $D$ meson. 
We determine the strong two--body decay widths $\sc \to \Lambda_c \pi$   
which are shown to be consistent with current data for the  
$J^{\rm P} = \frac{1}{2}^+$ and $J^{\rm P} = \frac{3}{2}^-$ assignments.  

\end{abstract}



\maketitle

A few years ago the Belle Collaboration~\cite{Mizuk:2004yu} observed 
an isotriplet of new baryon states with open charm $\Sigma_c(2800)$ 
decaying into $\Lambda_c \pi$. This resonance was fit by a D-wave
Breit-Wigner distribution based on the measured mass of the state
and the Belle Collaboration tentatively assigned the quantum 
numbers $J^P = \frac{3}{2}^-$. The same neutral state $\Sigma_c^0$ 
was possibly also observed in B decays by 
the {\it BABAR} Collaboration~\cite{Aubert:2008if}. Although the
measured width of this resonance is consistent with the Belle value, 
the mass value is higher and somewhat inconsistent with the previous 
measurement. The {\it BABAR} Collaboration indicates that there is
weak evidence that the excited $\Sigma_c^0$ they observe has $J =1/2$.
In the following we will assume that both collaborations observe the 
same baryon resonance, although the present mass discrepancy and the 
final assignment of quantum numbers remain to be resolved.

Earlier quark model predictions~\cite{Capstick:1986bm} for excited
baryons containing one charmed quark lead to a possible identification
of this state as a member of the $J^P = 3/2^-$ and $5/2^-$ doublet,
where the light quark subsystem carries total angular momentum of 2 units.
Further quark model studies on the mass spectrum of excited $\Sigma $ states
were also performed  
in~\cite{Garcilazo:2007eh,Ebert:2007nw,Gerasyuta:2007un,Roberts:2007ni}. 
These later works also tend to identify the $\Sigma_c(2800)$ as one of the  
nearly degenerate orbital excitations with $J^P = 1/2^-, 3/2^-$ or $5/2^-$. 
The strong decays of excited charmed baryons have been considered in the 
framework of heavy hadron chiral perturbation theory 
(HHChPT)~\cite{Cheng:2006dk} and in 
quark models~\cite{Chen:2007xf,Zhong:2007gp}.
The computed $\Lambda_c \pi$ decay widths~\cite{Chen:2007xf}
of various $J^P$ assignments for the $\Sigma_c(2800)$ in the context of 
the $^3P_0$ model are found to be inconsistent with observation. 
In the chiral model of~\cite{Zhong:2007gp} the $\Lambda_c \pi$ decay 
is consistent with a $J^P=1/2^-$ assignment, while the HHChPT framework 
of~\cite{Cheng:2006dk} uses the total width of $\Sigma_c(2800)$ as an input. 
Hence from the theoretical side a unique interpretation of this resonance 
in terms of an orbital excitation of the charmed three-quark system is 
presently not available.

Alternatively, in Ref.~\cite{Lutz:2005ip} it was suggested that the 
$\Sigma_c (2800)$ is a so-called chiral excitation of open-charm baryons 
with $J^P =3/2^-$ -- a charmed baryon resonance generated by S-wave 
coupled-channel dynamics of 
Goldstone bosons with the $J^P =3/2^+$ sextet baryons with open charm. 
When including further pseudoscalar mesons like $D$ mesons in the 
coupled-channel dynamics the updated results of~\cite{Hofmann:2006qx} 
resulted in $J^P =3/2^-$ state, where the width is much too large to 
justify the identification with the $\Sigma_c (2800)$.
The dynamical generation of charmed baryon resonances in the context of  
a unitarized meson-baryon coupled channel model was also pursued in 
Refs.~\cite{GarciaRecio:2008dp,JimenezTejero:2009vq}.
Both pseudoscalar and vector mesons are included in the S-wave coupled 
channel formalism as required by heavy quark symmetry. 
Now the $\Sigma_c (2800)$ is identified with a dynamically generated 
resonance in the $J^P =1/2^-$ channel with a dominant $ND$ configuration.
But the resulting partial $\Lambda_c \pi$  decay width is much too small 
to justify this identification.

Given the sparse experimental information concerning the excited $\Sigma_c$ 
spectrum and decays, but also the somewhat contradictory theoretical 
interpretations it is intruiging to note that the observed $\Sigma_c (2800)$ 
states are very close to the respective $ND$ thresholds. For example, the
measured mass difference $\Delta m = m_{\Sigma_c^0 (2800)} - m_{\Lambda_c^+} =
(515\pm 3 ^{+2}_{-6})$ MeV by the Belle Collaboration~\cite{Mizuk:2004yu}
corresponds to an absolute mass of $m_{\Sigma_c^0 (2800)}
= (2802^{+4}_{-7})$ MeV~\cite{Amsler:2008zz} which should be compared 
to the $nD^0$ threshold value of about 2804 MeV.
The closeness of these thresholds could imply that the $ND$ components
play a dominant role in the $\Sigma_c (2800)$ configurations, either
by coupling of the excited three-quark state to the $ND$ channels or
in a hadronic molecule configuration. 
Although a full dynamical calculation was not performed yet concerning 
binding in the $ND$ channel both for $J=1/2$ and $J=3/2$ (values for $J$ 
as suggested by experiment) here we pursue a possible hadronic molecule 
interpretation of the $\Sigma_c (2800)$
-- bound state of the nucleon and the charm $D$ meson.
Our aim is to work out in a hadronic framework the resulting
$\Sigma_c \to \Lambda_c \pi$ decay widths for possible quantum number
assignments of $J^P = 1/2^{\pm}$ and $3/2^{\pm}$ which will be
confronted with the experimental results.
Note that while $J^P =1/2^-$ corresponds to an S-wave $ND$ configuration,
the options $J^P = 1/2^+$ or $J^P = 3/2^+$ represent a P-wave
and $J^P = 3/2^-$ a relative D-wave in the $ND$ system.
Although slight binding in the $ND$ seems less likely for higher
partial waves, especially for the D-wave, the possibility of such a weakly
bound system is not excluded yet. 

In Refs.~\cite{Faessler:2007gv}-\cite{Branz:2009yt} 
we developed the formalism for the study of recently observed 
exotic meson states (like $D_{s0}^\ast(2317)$, $D_{s1}(2460)$, $X(3872)$, 
$Y(3940)$, $Y(4140$, $\cdots$) as hadronic molecules. 
The extension of our formalism to baryonic molecules 
has been done in Refs.~\cite{Dong:2008mt,Dong:2009tg}.  
A composite structure of these molecular states is defined  
by the compositeness condition $Z=0$~\cite{Weinberg:1962hj,%
Efimov:1993ei,Anikin:1995cf} 
(see also Refs.~\cite{Faessler:2007gv}-\cite{Dong:2009tg}).   
This condition implies that the renormalization constant of the hadron 
wave function is set equal to zero or that the hadron exists as a bound 
state of its constituents. The compositeness condition was originally 
applied to the study of the deuteron as a bound state of proton and
neutron~\cite{Weinberg:1962hj}. Then it was extensively used
in low--energy hadron phenomenology as the master equation for the
treatment of mesons and baryons as bound states of light and heavy
constituent quarks (see e.g. Refs.~\cite{Efimov:1993ei,Anikin:1995cf}). 
By constructing a phenomenological Lagrangian including the 
couplings of the bound state to its constituents and the constituents 
with other particles we calculated one--loop 
meson diagrams describing different decays of the molecular states 
(see details in~\cite{Faessler:2007gv}-\cite{Dong:2009tg}). 

In the present paper we proceed as follows. First, we discuss 
the basic notions of our approach. We consider a choice for 
the effective meson Lagrangian for the treatment of the 
$\Sigma_c$ baryons as $ND$ bound states: $\Sigma_c^{++} = (pD^+)$, 
$\Sigma_c^{+} = (pD^0 + nD^+)/\sqrt{2}$, $\Sigma_c^{0} = (nD^0)$, 
Second, we consider the two--body hadronic decays 
$\Sigma_c \to \Lambda_c + \pi$. 
Finally, we present our numerical results. 

We consider the triplet $(\Sigma_c^{++}, \Sigma_c^{+}, \Sigma_c^{0})$
as molecular states composed of nucleons and $D$ mesons as: 
\eq\label{Xstate}
|\Sigma_c^{++}\ra &=&   | p D^+ \ra  \, , \nonumber\\ 
|\Sigma_c^{+}\ra  &=&   \frac{1}{\sqrt{2}} 
| p D^0 + n D^+ \ra \, , \\
|\Sigma_c^{0}\ra &=&   | n D^0 \ra 
\nonumber 
\,. 
\en 
Our approach is based on an effective interaction Lagrangian describing 
the couplings of the $\sc$ to its constituents. 
The molecular structure of the $\sc$ baryon with quantum numbers 
$J^{\rm P} = \frac{1}{2}^\pm$ is described by the  Lagrangian
\eq\label{Lagr_Lc_S}
{\cal L}_{\sc}(x) &=& g_{_{\sc}} \, \overline{\bf \sc}(x)
\, {\bf J}_{\sc}(x) \,  + \ {\rm H.c.} \,, \hspace*{.5cm}
 {\bf J}_{\sc}(x) = D(x) \, {\bm\tau} \,
 \Gamma \, \int d^4y \, \Phi(y^2) \, N(x+y) 
\en
while for the choice $J^{\rm P} = \frac{3}{2}^\pm$ the Lagrangian contains
a derivative $ND$ coupling 
\eq\label{Lagr_Lc_D}
{\cal L}_{\sc}(x) &=& g_{_{\sc}} \, \overline{\bf \sc^\mu}(x)
\, {\bf J}_{\sc, \mu}(x) \,  + \ {\rm H.c.} \,, \hspace*{.5cm}
 {\bf J}_{\sc, \mu}(x) = D(x) \, {\bm\tau} \,
 \Gamma \, \int d^4y \, \Phi(y^2) \, \partial_\mu N(x+y) 
\en
where $g_{_{\sc}}$ is the coupling constant of the isotriplet $\sc$ to
the $N=( p ,n)^T$ and $D= (D^0 , D^+)^T$ constituents.
Here $\Gamma$ is the corresponding Dirac matrix
related to the spin--parity of the $\sc$.
In particular, for $J^{\rm P} = \frac{1}{2}^+ ,~ \frac{3}{2}^-$
we have $\Gamma = \gamma^5$ while for $J^{\rm P} = \frac{1}{2}^- , 
~\frac{3}{2}^+$ the Dirac structure
$\Gamma = I$ should be inserted in the $\sc$ current.

We propose a picture for 
the $\sc$ in analogy to heavy quark--light antiquark mesons, i.e. 
the heavy $D$ meson is located at the center of mass of the $\sc$, while 
the light nucleon surrounds the $D$. We describe the distribution of the 
nucleon around the $D$ meson by the correlation function 
$\Phi(y^2)$ depending on the relative Jacobi coordinate $y$. 
A basic requirement for the choice of an explicit form of the correlation 
function $\Phi(y^2)$ is that its Fourier transform vanishes sufficiently 
fast in the ultraviolet region of Euclidean space to render the Feynman 
diagrams ultraviolet finite. We adopt a Gaussian form for the correlation 
function. The Fourier transform of this function is given by
\eq 
\tilde\Phi(p_E^2/\Lambda^2) \doteq \exp( - p_E^2/\Lambda^2)\,,
\en 
where $p_{E}$ is the Euclidean Jacobi momentum. Here, 
$\Lambda \sim m_N \sim 1$ GeV is a size parameter, characterizing 
the distribution of the nucleon in the $\sc$ baryon, which is of order of 
the nucleon mass or 1 GeV. In the numerical analysis we therefore fix the mean
value to $\Lambda = 1$ GeV. 

The coupling constant $g_{_{\sc}}$ is determined by the 
compositeness condition~\cite{Weinberg:1962hj,Efimov:1993ei,Anikin:1995cf,%
Faessler:2007gv}. It implies that the renormalization constant of 
the hadron wave function is set equal to zero with:
\eq\label{ZLc}
Z_{\sc} = 1 - \Sigma_{\sc}^\prime(m_{\sc}) = 0 \,.
\en
Here, $\Sigma_{\sc}^\prime(m_{\sc}) = 
g_{_{\sc}}^2 \Pi^\prime_{\sc}(m_{\sc})$ is the
derivative of the mass operator for
$J = \frac{1}{2}$. For $J = \frac{3}{2}$ the same relation holds but now
$\Sigma_{\sc} (m_{\sc})$ should be identified with the scalar function
proportional to the Minkowski tensor $g^{\mu\nu}$
in the full mass operator $\Sigma^{\mu\nu}_{\sc}$.
Note, that for $J = \frac{3}{2}$ the other possible Lorentz structures 
in the $\sc$ mass operator vanish due to the Rarita--Schwinger conditions. 
The mass operator of the $\sc$ baryon is described by
the diagram of Fig.1. To clarify the physical meaning of the 
compositeness condition,
we first want to remind the reader that the renormalization constant
$Z_{\sc}^{1/2}$ can also be interpreted as the matrix element
between the physical and the corresponding bare state --- an elementary 
structureless field.
For $Z_{\sc}=0$ it then follows that the physical state
does not contain the bare one and hence is described as a bound state.
As a result of the interaction of the $\sc$ baryon with its constituents
$N$ and $D$, the $\sc$ baryon is dressed, i.e. its mass
and its wave function have to be renormalized. Note, in the present 
paper we only consider the contribution of a possible molecular $(ND)$ 
component to the structure of the $\sc$. An inclusion of a three--quark 
component is possible, but goes beyond the scope of the present paper. 

In Table I we display the results for the coupling $g_{_{\sc^+}}$ of the single
charged $\sc^+$ state for 
different spin--parity assignments and for a variation of the 
size parameter $\Lambda$ in the region of 0.75 -- 1.25 GeV. 
Note, that an increase of the $\Lambda$ value leads to an enhancement 
of the couplings $g_{_{\sc^+}}$.
A final value for the cutoff model parameter $\Lambda$ can ultimately only be 
fixed when more decay data on $\sc$ are available. 

\begin{figure}
\centering{\
\epsfig{figure=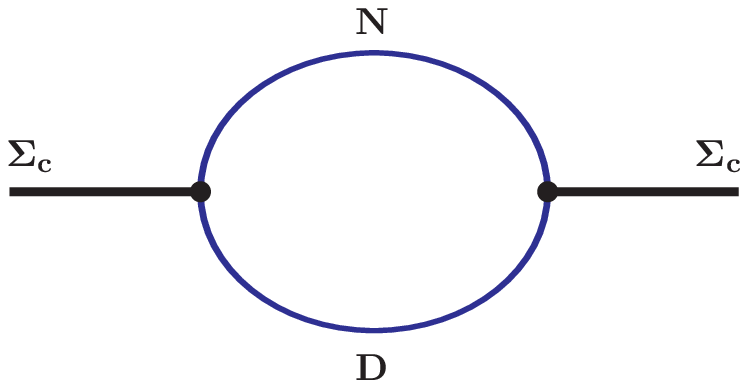,scale=.75}}
\caption{Diagram describing the $\Sigma_c$ mass operator.}
\label{fig:str}

\vspace*{1.5cm} 

\centering{\
\epsfig{figure=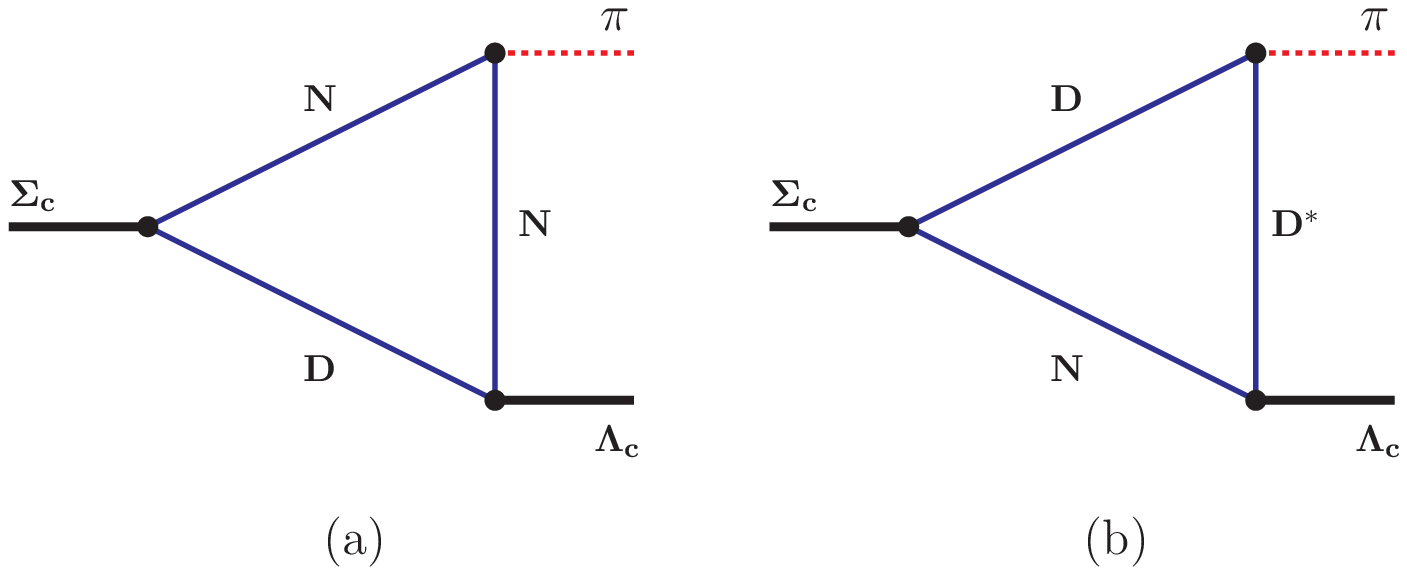,scale=.75}}
\caption{Diagrams contributing to the
$\Sigma_c \to \Lambda_c \pi$ decay.}
\label{fig:vertex}
\end{figure}

\begin{table}

\vspace*{1cm}

\begin{center}
{\bf Table I.} 
Coupling constants $g_{\sc^+}$ for different $J^P$ assignments. \\
Error reflects variation in $\Lambda$ from 0.75 to 1.25 GeV. 
\vspace*{.25cm}

\def\arraystretch{1.5}
\begin{tabular}{|c|c|c|c|}  
\hline 
          \ \ \ $J^{\rm P} = \frac{1}{2}^+$ \ \ \ 
        & \ \ \ $J^{\rm P} = \frac{1}{2}^-$ \ \ \ 
        & $J^{\rm P} = \frac{3}{2}^+$ 
        & $J^{\rm P} = \frac{3}{2}^-$ \\ 
\hline
7 $\pm$ 1.9 & 0.6 $\pm$ 0.2 & 
4.2 $\pm$ 1.4 GeV$^{-1}$ & 35.3 $\pm$ 1.8 GeV$^{-1}$ \\
\hline
\end{tabular}
\end{center}
\end{table} 

The one--loop hadron diagrams contributing to the $\sc \to \lc + \pi$
transition are displayed in Fig.2(a) and 2(b). 
To evaluate these strong $\sc \to \lc + \pi$ decays 
we further need an effective Lagrangian including
the couplings of the $\sc$ constituents to $\lc$ and $\pi$. 
The $\pi NN$ and $D^\ast D \pi$ couplings are constrained by 
data or by low--energy theorems. In particular, 
the $\pi NN$ coupling at leading order of the chiral expansion is expressed 
through the nucleon mass $m_N$, the pion decay constant $F_\pi = 92.4$ MeV 
and the nucleon axial charge $g_A = 1.2695$ as: 
\eq
{\cal L}_{\pi NN} = g_{\pi NN} \bar N \, 
i \gamma_5 \bm{\pi} \, \bm{\tau} \, N\,, 
\hspace*{.5cm} g_{\pi NN} = \frac{m_N}{F_\pi} \, g_A \, .  
\en 

For the $D^\ast D \pi$ coupling we take the 
central value of $g_{D^\ast D \pi} = 17.9$ extracted from 
the measured $D^{\ast +} \to D^0 + \pi^+$ decay width~\cite{Anastassov:2001cw} 
\eq 
\Gamma(D^{\ast +} \to D^0 \pi^+) = 
\frac{g_{D^\ast D \pi}^2}{24 \pi m_{D^{\ast +}}^2} \, P^{\ast \, 3} \,, 
\en 
where $P^\ast$ is the three--momentum of $\pi^+$ in 
the $D^{\ast +}$ rest frame.  
Then the interaction $D^\ast D \pi$ Lagrangian reads:  
\eq 
{\cal L}_{D^\ast D \pi} = 
\frac{g_{D^\ast D \pi}}{\sqrt{2}} \, 
D^{\ast\, \dagger}_\mu \, i \partial^\mu \bm{\pi} \, \bm{\tau} D 
\ + \ {\rm H.c.} 
\en 
The couplings $D N \lc$ and $D^\ast N \lc$ are estimated 
by matching the flavor SU(4) effective Lagrangian to the SU(3) version, 
both describing the couplings of pseudoscalar 
and vector mesons to two baryons (see details in the Appendix): 
\eq 
g_{DN\Lambda_c} = - g_{\pi NN}\,, \hspace*{.5cm} 
g_{D^\ast N\Lambda_c} = - \frac{\sqrt{3}}{2} \, 
g_{\rho NN}\, . \hspace*{.5cm}
\en  
In the evaluation of the diagrams in Figs.1 and 2 we use the
standard free propagators for the intermediate particles: 
\eq 
iS_N(x-y)=\left<0|TN(x)\bar N(y)|0\right>=\int\frac{d^4k}{(2\pi)^4i}\, 
e^{-ik(x-y)} S_N(k),\quad S_N(k)=
\frac{1}{m_N - \not\! k-i\epsilon}
\en 
for nucleons, 
\eq
iS_{D}(x-y)=
\left<0|TD(x)D^{\dagger} (y)|0\right>=
\int\frac{d^4k}{(2\pi)^4i}\,e^{-ik(x-y)} S_D(k)\,,
\quad S_D(k)=
\frac{1}{m_D^2-k^2-i\epsilon}
\en 
for pseudoscalar $D$ mesons and 
\eq
iS_{D^\ast}^{\mu\nu}(x-y)=
\left<0|TD^{\ast\,\mu}(x)D^{\ast\, \nu \dagger}(y)|0\right>=
\int\frac{d^4k}{(2\pi)^4i}\,e^{-ik(x-y)} S^{\mu\nu}_{D^\ast}(k)\,,
\quad S_{D^\ast}^{\mu\nu}(k)=\frac{-g^{\mu\nu} + 
k^\mu k^\nu/m_{D^\ast}^2}{m_{D^\ast}^2-k^2-i\epsilon}
\en 
for vector $D^\ast$ mesons. 

The decay widths of the strong two--body transitions $\sc \to \lc + \pi$ 
are then calculated according to the following expressions for the 
different spin--parity assignments $\sc[J^P]$: 
\seq\label{Gamma}  
\eq
\Gamma(\sc[1/2^+] \to \lc + \pi) &=& 
\frac{d_{\sc\lc\pi}^2}{16\pi m_{\sc}^3} \, 
\lambda^{1/2}(m_{\sc}^2,m_{\lc}^2,m_\pi^2) 
\Big( (m_{\sc} - m_{\lc})^2 - m_\pi^2 \Big) \,, \\
\Gamma(\sc[1/2^-] \to \lc + \pi) &=& 
\frac{h_{\sc\lc\pi}^2}{16\pi m_{\sc}^3} \, 
\lambda^{1/2}(m_{\sc}^2,m_{\lc}^2,m_\pi^2) 
\Big( (m_{\sc} + m_{\lc})^2 - m_\pi^2 \Big) \,, \\
\Gamma(\sc[3/2^+] \to \lc + \pi) &=& 
\frac{f_{\sc\lc\pi}^2}{192\pi m_{\sc}^5} \, 
\lambda^{3/2}(m_{\sc}^2,m_{\lc}^2,m_\pi^2) 
\Big( (m_{\sc} - m_{\lc})^2 - m_\pi^2 \Big) \,, \\
\Gamma(\sc[3/2^-] \to \lc + \pi) &=& 
\frac{g_{\sc\lc\pi}^2}{192\pi m_{\sc}^5} \, 
\lambda^{3/2}(m_{\sc}^2,m_{\lc}^2,m_\pi^2) 
\Big( (m_{\sc} + m_{\lc})^2 - m_\pi^2 \Big) \,,  
\en 
\sen 
where $\lambda(x,y,z) = x^2 + y^2 + z^2 - 2xy - 2yz - 2xz$ is 
the K\"allen function; $m_{\sc}$, $m_{\lc}$ and $m_\pi$ are the masses 
of $\sc$, $\lc$ baryons and the pion. We also introduce the effective
coupling constants $d_{\sc\lc\pi}$, $h_{\sc\lc\pi}$ 
$f_{\sc\lc\pi}$ and $g_{\sc\lc\pi}$ defining the interaction of
the $\sc$ of definite spin--parity with $\lc$ and $\pi$ 
as a result of the processes in Fig.2 with: 
\seq
\eq 
{\cal L}_{\sc(1/2^+)\lc\pi} &=& d_{\sc\lc\pi} 
\overline\lc \, \gamma_5 \, \bm{\pi} \, 
\bm{\Sigma}_c + {\rm H.c.}  \,, \\
{\cal L}_{\sc(1/2^-)\lc\pi} &=& h_{\sc\lc\pi} 
\overline\lc \, \bm{\pi} \, 
\bm{\Sigma}_c + {\rm H.c.} \,, \\
{\cal L}_{\sc(3/2^+)\lc\pi} &=& f_{\sc\lc\pi} 
\overline\lc \, \partial_\mu \bm{\pi} \, 
\bm{\Sigma}_c^{\mu} + {\rm H.c.}  \,,  \\ 
{\cal L}_{\sc(3/2^-)\lc\pi} &=& g_{\sc\lc\pi} 
\overline\lc \, \gamma_5 \partial_\mu \bm{\pi} \, 
\bm{\Sigma}_c^{\mu} + {\rm H.c.} \,. 
\en 
\sen 
In Table II we present our results for 
these effective couplings $d_{\sc\lc\pi}$, $h_{\sc\lc\pi}$, 
$f_{\sc\lc\pi}$ and $g_{\sc\lc\pi}$ including a variation of the 
cutoff parameter from 0.75 to 1.25 GeV.   

\begin{table}
\begin{center}
{\bf Table II.} 
Effective couplings $d_{\sc\lc\pi}$, 
$h_{\sc\lc\pi}$, $f_{\sc\lc\pi}$ and $g_{\sc\lc\pi}$.  \\
Error reflects variation in $\Lambda$ from 0.75 to 1.25 GeV.  

\vspace*{.25cm}

\def\arraystretch{1.5}
\begin{tabular}{|c|c|c|c|c|}  
\hline 
Mode & $d_{\sc\lc\pi}$  
     & $h_{\sc\lc\pi}$  
     & $f_{\sc\lc\pi}$  
     & $g_{\sc\lc\pi}$  \\
\hline
$\Sigma_c^{++} \to \Lambda_c^+ \pi^+$ 
& $-$ 8.15 $\pm$ 2.72  
& 1.63 $\pm$ 0.54 
& 1.95 $\pm$ 0.97 GeV$^{-1}$ 
& $-$ 3.35 $\pm$ 1.61 GeV$^{-1}$ \\ 
\hline 
$\Sigma_c^{+} \to \Lambda_c^+ \pi^0$
& $-$ 7.78 $\pm$ 2.60 
& 1.48 $\pm$ 0.47  
& 1.90 $\pm$ 0.95 GeV$^{-1}$ 
& $-$ 3.24 $\pm$ 1.59 GeV$^{-1}$ \\ 
\hline 
$\Sigma_c^{0} \to \Lambda_c^+ \pi^-$ 
& $-$ 7.52 $\pm$ 2.54  
& 1.43 $\pm$ 0.45  
& 1.87 $\pm$ 0.94 GeV$^{-1}$  
& $-$ 3.16 $\pm$ 1.58 GeV$^{-1}$ \\
\hline
\end{tabular}
\end{center}
\end{table} 

\begin{table}
\begin{center}
{\bf Table III.} 
$\sc \to \lc\pi$ decay widths (in MeV) 
for different spin--parity assignments of the $\sc$. \\
Error reflects variation in $\Lambda$ from 0.75 to 1.25 GeV. 
Results for preferred value of $\Lambda = 1$ GeV are given in brackets. 

\vspace*{.25cm}

\def\arraystretch{1.5}
\begin{tabular}{|c|c|c|c|c|}  
\hline 
Mode & $J^{\rm P} = \frac{1}{2}^+$ 
     & $J^{\rm P} = \frac{1}{2}^-$ 
     & $J^{\rm P} = \frac{3}{2}^+$ 
     & $J^{\rm P} = \frac{3}{2}^-$ \\ 
\hline
$\Sigma_c^{++} \to \Lambda_c^+ \pi^+$ 
& 41.1 $\pm$ 24.7 (37.0) 
& 173.6 $\pm$ 103.6 (156.4) 
& 0.176 $\pm$ 0.140 (0.141) 
& 54.5 $\pm$ 42.6 (44.3)  \\
\hline 
$\Sigma_c^{+} \to \Lambda_c^+ \pi^0$
& 37.6 $\pm$ 22.6 (33.9) 
& 142.3 $\pm$ 82.1 (129.3) 
& 0.171 $\pm$ 0.137 (0.137)  
& 51.9 $\pm$ 41.0 (41.8) \\
\hline 
$\Sigma_c^{0} \to \Lambda_c^+ \pi^-$
& 35.1 $\pm$ 21.3 (31.5) 
& 132.3 $\pm$ 75.8 (120.4) 
& 0.164 $\pm$ 0.132 (0.131)
& 49.5 $\pm$ 39.6 (39.6) \\ 
\hline
\end{tabular}
\end{center}
\end{table} 

Our final numerical results for the decay widths 
are summarized in Table III. For the $\sc$ masses 
we use the measured values~\cite{Mizuk:2004yu} of the Belle Collaboration. 
The predictions for the decay widths differ sizably 
depending on the $J^P$ assignment for the $\sc(2800)$.
These predictions are to be compared to the measured total widths of the 
$\sc (2800)$ baryons~\cite{Mizuk:2004yu,Amsler:2008zz} with:
\eq
& &\Gamma(\Sigma_c^{++}) = 75^{+ 18 + 12}_{- 13 - 11} \ {\rm MeV}\,,
\nonumber\\
& &\Gamma(\Sigma_c^{+}) = 62^{+ 37 + 52}_{- 23 - 38} \ {\rm MeV}\,,
\\
& &\Gamma(\Sigma_c^{0}) = 61^{+ 18 + 22}_{- 13 - 13} \ {\rm MeV}\, .
\nonumber
\en
Since the observed $\Lambda_c \pi$ decay modes of the $\sc(2800)$ 
states are assumed to be dominant present results favor, at least in 
the context of the $ND$ molecule interpretation, either the 
$J^P = 1/2^+$ or the $J^P = 3/2^-$ assignment. Note, the $J^P = 3/2^-$   
assignment was originally assumed by the Belle 
Collaboration~\cite{Mizuk:2004yu}. 
The alternative scenario for $\sc$ with $J^P = 3/2^+$ 
is clearly excluded by the predictions of Table III. 
For small values of the dimensional parameter $\Lambda$ 
the scenario for $\sc$ with $J^P = 1/2^-$ becomes compatible 
with data. 

In conclusion, we estimated the strong $\Lambda_c \pi$ decays of the 
$\sc(2800)$ baryon for different spin--parity assignments 
assuming a dominant molecular $ND$ structure of 
this state. Judging from the decay widths of the order of 40 MeV
we find that the original scenario where the $\sc$ has spin--parity 
$J^{\rm P} = \frac{3}{2}^-$ and the choice $J^{\rm P} = \frac{1}{2}^+$ 
are consistent with current data.
The option $J^{\rm P} = \frac{3}{2}^+$ leads to strongly  
suppressed partial decay widths of the order of a hundred keV,  
while $J^{\rm P} = \frac{1}{2}^-$ leads to enhanced
partial decay widths and only becomes compatible with data 
for relatively small values of the dimensional parameter $\Lambda$.  
 Although weak binding in the 
$ND$ system for $J^P =1/2^+$ and $J^P =3/2^-$ remains 
to be studied, present evaluation of the $\Lambda_c \pi$ decay widths 
point to a possibly sizable role of the $ND$ configuration in the $\sc(2800)$.

\newpage 
 
\begin{acknowledgments}

This work was supported by the DFG under Contract No. FA67/31-2 
and No. GRK683. This work is supported by the National Sciences Foundations 
No. 10775148 and 10975146 and by CAS grant No. KJCX3-SYW-N2 (YBD). 
This research is also part of the
European Community-Research Infrastructure Integrating Activity
``Study of Strongly Interacting Matter'' (HadronPhysics2,
Grant Agreement No. 227431), Russian President grant
``Scientific Schools''  No. 3400.2010.2, Russian Science and
Innovations Federal Agency contract No. 02.740.11.0238. 

\end{acknowledgments}

\appendix\section{Matching of the phenomenological SU(3) and SU(4)
$PBB$ and $VBB$ interaction Lagrangians} 

First we consider the $PBB$ interaction that is the coupling of a
pseudoscalar ($P$) meson to two baryons ($BB$). 
In flavor SU(3) the couplings are generated
from the ${\cal O}(p)$ term of 
chiral perturbation theory (ChPT)~\cite{ChPT} describing 
the coupling of baryon fields with the chiral fields: 
\eq 
{\cal L}_{PBB}^{SU_3} = -\frac{D}{F_P\sqrt{2}} (m_B + m_{\bar B}) 
{\rm tr} \Big(\bar B i \gamma^5 \{ P B \}\Big) 
 \, - \, \frac{F}{F_P\sqrt{2}} (m_B + m_{\bar B}) {\rm tr} 
\Big(\bar B i \gamma^5  [ P B ]\Big) \, .
\en 
Here $F_P = F_\pi = 92.4$ MeV is the 
leptonic decay constant;  $D$ and $F$ are the baryon axial coupling constants 
(we restrict to the SU(3) symmetric limit, where $D = 3F/2 = 3g_A/5$ 
with $g_A = 1.2695$ being the nucleon axial charge) 
the symbols ${\rm tr}$, $\{ \ldots \}$ and $[ \ldots ]$ denote the trace 
over flavor matrices, anticommutator and commutator, respectively.
We replace the pseudovector coupling by the pseudoscalar one 
considering on-mass-shell baryons. 
The SU(3) baryon $B$ and pseudoscalar meson $P$
matrices read as: 
\eq 
B = 
\left(
\begin{array}{ccc}
\Sigma^0/\sqrt{2} + \Lambda/\sqrt{6}\,\, & \,\, \Sigma^+ \,\, & \, p \\
\Sigma^- \,\, & \,\, -\Sigma^0/\sqrt{2}+\Lambda/\sqrt{6}\,\, & \,  n\\
\Xi^-\,\, & \,\, \Xi^0 \,\, & \, -2\Lambda/\sqrt{6}\\
\end{array}
\right),  
\en 
\eq 
P = 
\left(
\begin{array}{ccc}
\pi^0/\sqrt{2} + \eta/\sqrt{6}\,\, & \,\, \pi^+ \,\, & \, K^+ \\
\pi^- \,\, & \,\, -\pi^0/\sqrt{2}+\eta/\sqrt{6}\,\, & \, K^0\\
K^-\,\, & \,\, \bar K^0 \,\, & \, -2\eta/\sqrt{6}\\
\end{array}
\right) \, .  
\en 
The SU(4) $PBB$ Lagrangian is given by~\cite{Okubo:1975sc}: 
\eq 
{\cal L}_{PBB}^{SU_4} = g_1 \bar B^{kmn} i\gamma_5 P^l_k B_{lmn} 
                      + g_2 \bar B^{kmn} i\gamma_5 P^l_k B_{lnm} \, ,
\en 
where the indices $l,m,n$ of the tensor $B_{lmn}$
run from 1 to 4, representing the 20--plet of baryons 
(see details in Refs.~\cite{Okubo:1975sc});
$P^l_k$ is the matrix representing the 15--plet of
pseudoscalar fields. The baryon tensor satisfies the conditions 
\eq 
B_{lmn} + B_{mnl} + B_{nlm} = 0, \hspace*{.5cm} 
B_{lmn} = B_{mln} \,. 
\en 
The full list of physical states in terms of SU(4) tensors is 
given in Ref.~\cite{Okubo:1975sc}. Here we only display a few of them: 
\eq 
& &p = B_{112} = - 2 B_{121} = - 2 B_{211}\,, \hspace*{.25cm}
   n = - B_{221} = 2 B_{212} = 2 B_{122}\,, \nonumber\\ 
& &\Sigma_c^{++} = B_{114} = - 2 B_{141} = - 2 B_{411}\,, \hspace*{.25cm}
   \Sigma_c^0    = - B_{224} = 2 B_{242} = 2 B_{422}\,, \\ 
& &\pi^+ = P^2_1\,, \hspace*{.25cm} 
   \pi^- = P^1_2\,, \hspace*{.25cm}  
    D^0  = P^1_4\,, \hspace*{.25cm} 
    D^{\ast +} = V^2_4\,, \hspace*{.25cm}  
    D^{\ast 0} = V^1_4 \,. \nonumber 
\en  
Evaluating the $\pi NN$ couplings in both versions we fix 
the SU(4) couplings $g_1$ and $g_2$ as (in the SU(3) Lagrangian 
we restrict to the mass degenerate case $m_B = m_{\bar B} 
= m_p = 938.27$ MeV):  
\eq 
g_{\pi NN} = g_1 - \frac{5}{4} g_2\,, \hspace*{.25cm} 
g_{\pi NN} \frac{D-F}{D+F}= - \frac{g_1 + g_2}{4 \sqrt{2}}.  
\en 
Considering the SU(3) symmetric ratio of $F$ and $D$ couplings 
$F/D = 2/3$ we get 
\eq 
g_1 = 0\,, \hspace*{.25cm} 
g_2 = - \frac{4}{5} \sqrt{2} \, g_{\pi NN} \,. 
\en 
Finally the $g_{DN\lc}$ coupling is fixed as 
\eq 
g_{DN\Lambda_c} = - g_{\pi NN}\,. 
\en  
In complete analogy we fix the vector meson $VBB$ couplings. 
The SU(3) $VBB$ Lagrangian 
can be expressed in terms of the $\rho NN$ coupling constant as: 
\eq 
{\cal L}_{VBB}^{SU_3} = \frac{g_{\rho NN}}{\sqrt{2}} 
{\rm tr} \Big(\bar B \gamma^\mu \{ V_\mu B \}\Big) 
\, + \, \frac{g_{\rho NN}}{\sqrt{2}} {\rm tr} 
\Big(\bar B \gamma^\mu  B\Big) {\rm tr}V_\mu \,
\en 
where 
\eq 
V = 
\left(
\begin{array}{ccc}
\rho^0/\sqrt{2} + \omega/\sqrt{2}\,\, & \,\, \rho^+ \,\, & \, K^{\ast +} \\
\rho^- \,\, & \,\, -\rho^0/\sqrt{2}+\omega/\sqrt{2}\,\, & \, K^{\ast 0}\\
K^{\ast -}\,\, & \,\, \bar K^{\ast 0} \,\, & \, - \phi \\
\end{array}
\right). 
\en 
The SU(4) $VBB$ Lagrangian is given by~\cite{Okubo:1975sc}: 
\eq 
{\cal L}_{PBB}^{SU_4} = h_1 \bar B^{kmn} \gamma^\mu V^l_{\mu, k} B_{lmn} 
                      + h_2 \bar B^{kmn} \gamma^\mu V^l_{\mu, k} B_{lnm} \, .
\en 
Evaluating the $\rho NN$ couplings in both versions we fix 
the SU(4) couplings $h_1$ and $h_2$ as: 
\eq 
h_1 = 2 h_2 = \frac{8}{3\sqrt{2}} g_{\rho NN} \, .
\en
Finally, the $D^\ast N\lc$ coupling is fixed as 
\eq 
g_{D^\ast N\Lambda_c} = - \frac{\sqrt{3}}{2} \, 
g_{\rho NN}\, ,
\en  
where for the $g_{\rho NN}$ coupling we take the SU(3) prediction of
\eq
g_{\rho NN} = 6 \, .
\en

\end{document}